# Age-Stratified Differences in Morphological Connectivity Patterns in ASD: An sMRI and Machine Learning Approach

Gokul Manoj[1*], Sandeep Singh Sengar[2], Jac Fredo Agastinose Ronickom[1]
[1]*School of Biomedical Engineering, Indian Institute of Technology (BHU), Varanasi, Uttar Pradesh, India.*
[2]*Department of Computer Science, Cardiff Metropolitan University, Cardiff, United Kingdom.*

[*]*Corresponding Author: Email: gokulmanoj.bme18@itbhu.ac.in*

## Abstract

**Purpose**: Age biases have been identified as an essential factor in the diagnosis of ASD. The objective of this study was to compare the effect of different age groups in classifying ASD using morphological features (MF) and morphological connectivity features (MCF).

**Methods:** The structural magnetic resonance imaging (sMRI) data for the study was obtained from the two publicly available databases, ABIDE-I and ABIDE-II. We considered three age groups, 6 to 11, 11 to 18, and 6 to 18, for our analysis. The sMRI data was pre-processed using a standard pipeline and was then parcellated into 148 different regions according to the Destrieux atlas. The area, thickness, volume, and mean curvature information was then extracted for each region which was used to create a total of 592 MF and 10,878 MCF for each subject. Significant features were identified using a statistical t-test ($p$<0.05) which was then used to train a random forest (RF) classifier.

**Results:** The results of our study suggested that the performance of the 6 to 11 age group was the highest, followed by the 6 to 18 and 11 to 18 ages in both MF and MCF. Overall, the MCF with RF in the 6 to 11 age group performed better in the classification than the other groups and produced an accuracy, F1 score, recall, and precision of 75.8%, 83.1%, 86%, and 80.4%, respectively.

**Conclusion:** Our study thus demonstrates that morphological connectivity and age-related diagnostic model could be an effective approach to discriminating ASD.

## 1. Introduction

Autism spectrum disorder (ASD) is a neurodevelopmental condition characterized by deficits in social communication and interaction, as well as restricted, repetitive patterns of behavior, interests, or activities [1]. It is a heterogeneous abnormality that causes altered cortical anatomy, abnormal white matter integrity, and altered brain function [2]. The underlying neural mechanisms of ASD are poorly understood, and its diagnosis mainly relies upon subjective evaluation and may result in prolonged or misdiagnosis of the condition [3]. Studies have shown that the morphological changes in the brain can be used as an effective biomarker for the diagnosis of ASD [4]. Structural magnetic resonance imaging (sMRI) is a widely used technique to study these anatomical variations of the ASD brain [5]. Numerous investigations have utilized univariate analysis techniques, specifically focusing on voxel-wise or local morphological features (MF) such as surface area, thickness, volume, and mean curvature of distinct brain areas, in order to analyze the brain of individuals with ASD utilizing sMRI images [6]. Nevertheless, these methodologies are inadequate in terms of identifying the inter-regional correlations among distinct brain regions. Morphological connectivity features (MCF) provide a method of obtaining higher-order cortical information related to brain areas by examining interregional morphological correlations between pairs of regions. This analytical approach holds potential as a significant diagnostic tool for ASD [7][8]. It can provide insights into the underlying neural mechanisms behind both brain function and dysfunction, and studies have also demonstrated the importance of MCF in classifying ASD and proved that it outperformed MF [9].

Machine learning algorithms trained on these anatomical features could be useful for studying ASD [10]. Studies have reported random forest (RF) algorithm as an optimal classifier for the smaller sample size [11]. Moreover, our past study has also highlighted the effectiveness of using RF over other classifiers [9]. However, the development of a unified classification model for the diagnosis of ASD is complicated due to the highly heterogeneous nature of ASD. Past studies have shown that an age-stratified approach to identifying the characteristics of ASD could be an effective method to mitigate the heterogeneity present in the condition [12], [13]. In this study, we attempted to compare the performance of RF classifier in different age groups of ASD and typical developing (TD) subjects using various MF and MCF obtained from sMRI. By exploring the performance of the classifier in different age strata, the study sought to gain insights into potential age-related differences in brain connectivity patterns and morphological characteristics associated with ASD.

## 2. Methods

### 2.1. Database

We considered a total of 313 ASD and 397 TD participants obtained from the 7 sites of the two open-access databases, Autism Brain Imaging Data Exchange (ABIDE-I and ABIDE-II) for our study [14], [15] The ABIDE database contains a collection of the sMRI and corresponding resting-state functional MRI and phenotypic information from over 17 different sites with the participant's demographic information and diagnostic status. Detailed demographic information of the 710 subjects is given in Table 1.

**Table 1** Demographic information of the subjects

| | **6 to 11 years** | | **11 to 18 years** | | **6 to 18 years** | |
|---|---|---|---|---|---|---|
| | **TD** | **ASD** | **TD** | **ASD** | **TD** | **ASD** |
| Count | 177 | 129 | 220 | 184 | 397 | 313 |
| Gender | 119 M 58 F | 106 M 23 F | 175 M 45 F | 167 M 17 F | 294 M 103 F | 273 M 40 F |
| FIQ/PIQ (Mean ± SD) | 117.2 ± 12.1 | 105.3 ± 18.0 | 111.6 ± 13.0 | 104.6 ± 15.6 | 114.1 ± 12.9 | 104.9 ± 16.6 |

**M:** Male**; F:** Female**; SD:** Standard deviation**; FIQ:** Full-scale intelligence quotient; **PIQ:** Performance intelligence quotient

### 2.2. Process Pipeline

Figure 1 represents the process pipeline for the age-stratified analysis of ASD subjects. It involves the following steps: 1) Segregation of the sMRI data into three different age groups, 2) Pre-processing of the sMRI data, 3) Extraction of MF and MCF features, 4) Classification using RF classifier and analysis of brain networks.

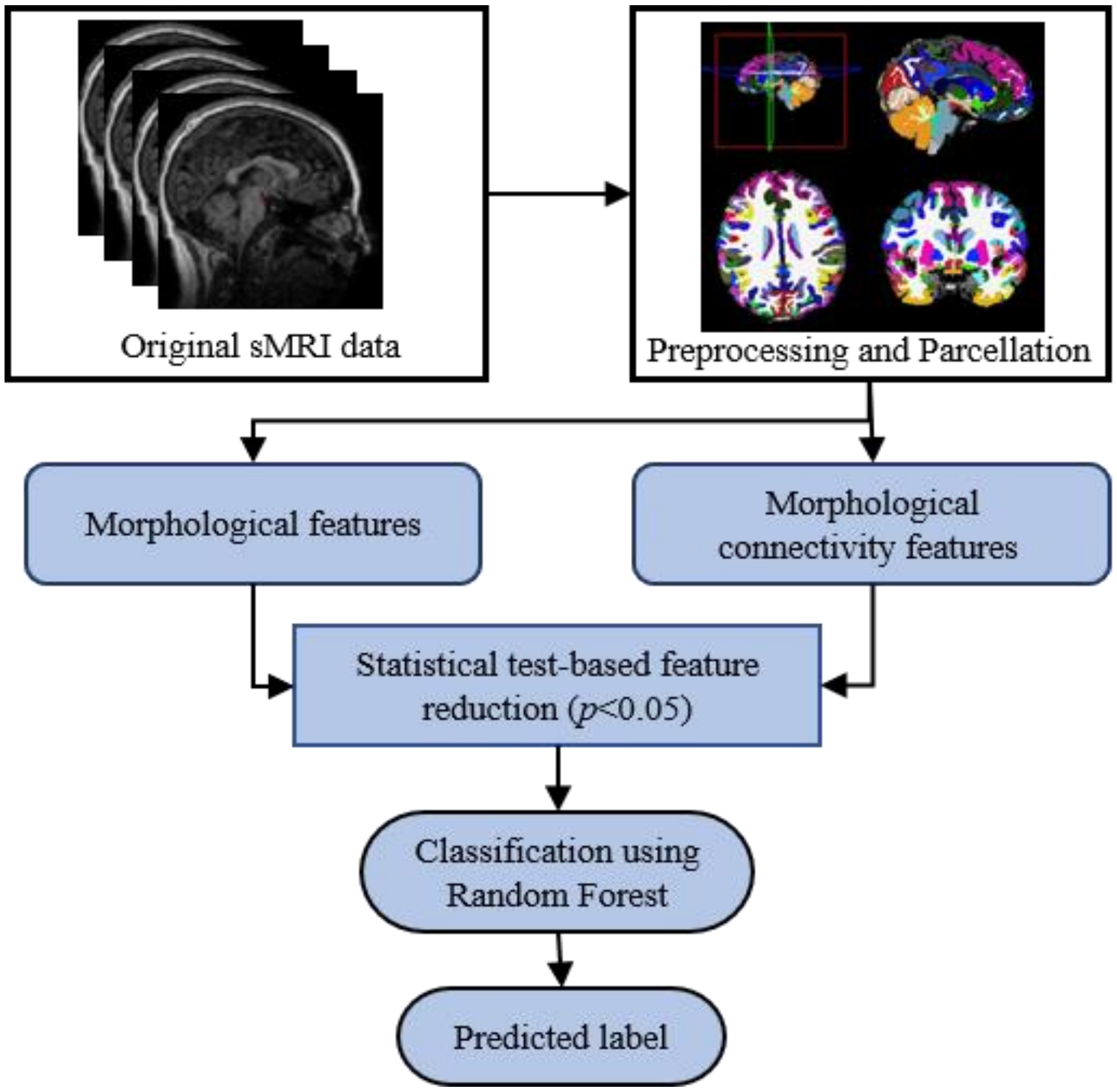

**Fig 1**. Process pipeline used for the study

### 2.3. Pre-processing

All the subjects were divided into three age groups, 6 to 11, 11 to 18, and 6 to 18, for the analysis. The sMRI images were pre-processed using the FreeSurfer toolbox [16]. The pre-processing pipeline included three stages of recon: autorecon1, autorecon2, and autorecon3. Overall, the process involved motion correction, intensity normalization, skull stripping, labeling based on Gaussian classifier atlas models, white matter segmentation, and cortical and white matter parcellation. We used the Destrieux atlas for the parcellation, and the brain regions were divided into 148 different segments (74 from each hemisphere).

### 2.4. Feature Extraction

From the segmented data, we extracted the surface area, thickness, volume, and mean curvature. The data was then standardized, and the MF was computed by combining the four measures to produce 592 (148x4) features. To calculate the MCF, 1D arrays were created using the four measures for each region. We estimated the MCF for the different regions using the Euclidean distance measure. It is computed as:

$$d(a,b) = \sqrt{[\Sigma(a_i - b_i)^2]}$$

Where $a$ and $b$ are two regions, and $i$ is the specific measure. For each subject, we obtained 10,878 features (148 x (148-1)/2), and statistical test-based ($p$-value) feature reduction was performed to reduce the number of features in both cases. A two-sample t-test was performed on them to calculate the $p$-value for different features. Features that had a $p$-value less than 0.05 were selected in each case to create the final feature set.

### 2.5. Classification

The feature set corresponding to the different age groups was then used to train the RF classifier with default parameters and a train-test split of 80% and 20%, respectively. We evaluated the performance of the model using the accuracy, recall, precision, and F1 score. RF is one of the popular machine-learning algorithms for ASD diagnosis. As an ensemble learning method, RF constructs a set of decision trees during the training phase, where each tree is trained on a random subset of the data. The final prediction is then made by aggregating the outputs of all individual trees. Notably, the RF classifier employs random feature selection during the decision tree construction, ensuring only a random subset of features is considered at each node for splitting. This feature randomness mitigates the risk of overfitting, enhancing the classifier's robustness and generalizability to new data [17].

## 3. Results

A raw sample of the T1-weighted sMRI obtained from the ABIDE database is shown in Figure 2 (a). Its corresponding skull-stripped image and Destrieux atlas parcellation in three different views- trans axial, sagittal, and coronal, are shown in Figure 2 (b) and (c), respectively. The images were also visually inspected to ensure that the skull-stripping and parcellation were performed with high quality.

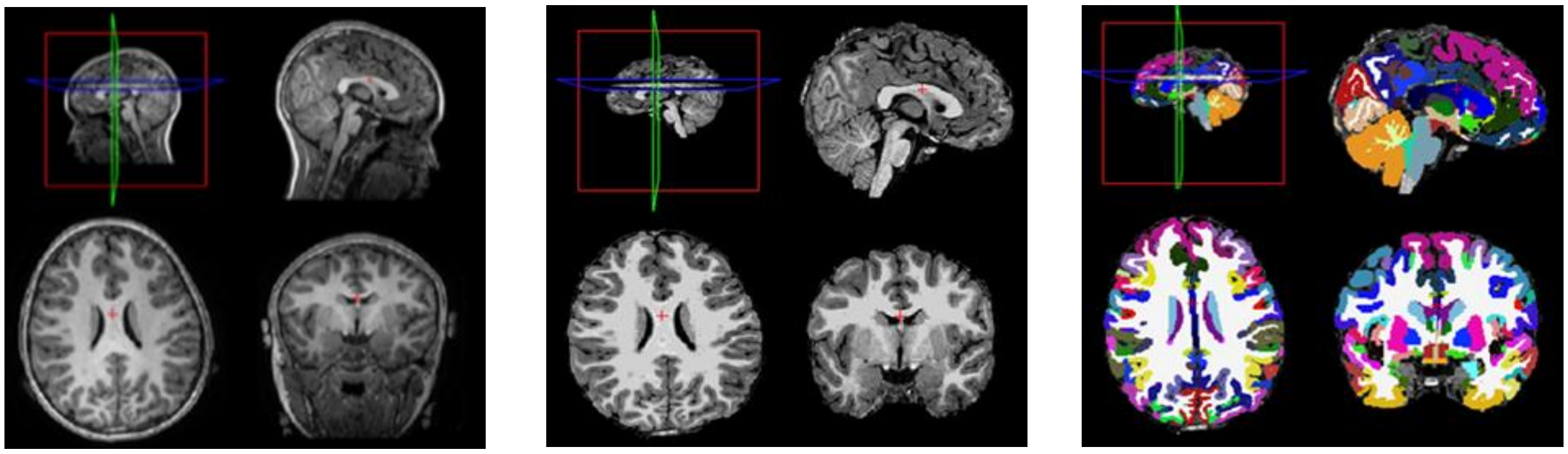

(a) (b) (c)

**Fig. 2** (a) 3D sMRI image, (b) Skull-stripped image and (c) Destrieux atlas parcellations

We computed the surface area, thickness, volume, and mean curvature of each ROI to obtain the MF and MCF. The features were then ranked using a statistical test, and the features that had a $p$- value less than 0.05 were selected in both MF and MCF. These features were then used to train an RF classifier, and we compared the age-specific performance of the classifier on the three age groups; 6 to 11, 11 to 18, and 6 to 18. The results of the classification are given in Figure 3. It was observed that within each type of analysis, the accuracy of the 6 to 11 age group was the highest (67.7% for MF and 75.8% for MCF), followed by the 6 to 18 (60.36% for MF and 67.6% for MCF) and 11 to 18 (59.3% for MF and 56.8% for MCF) age group.

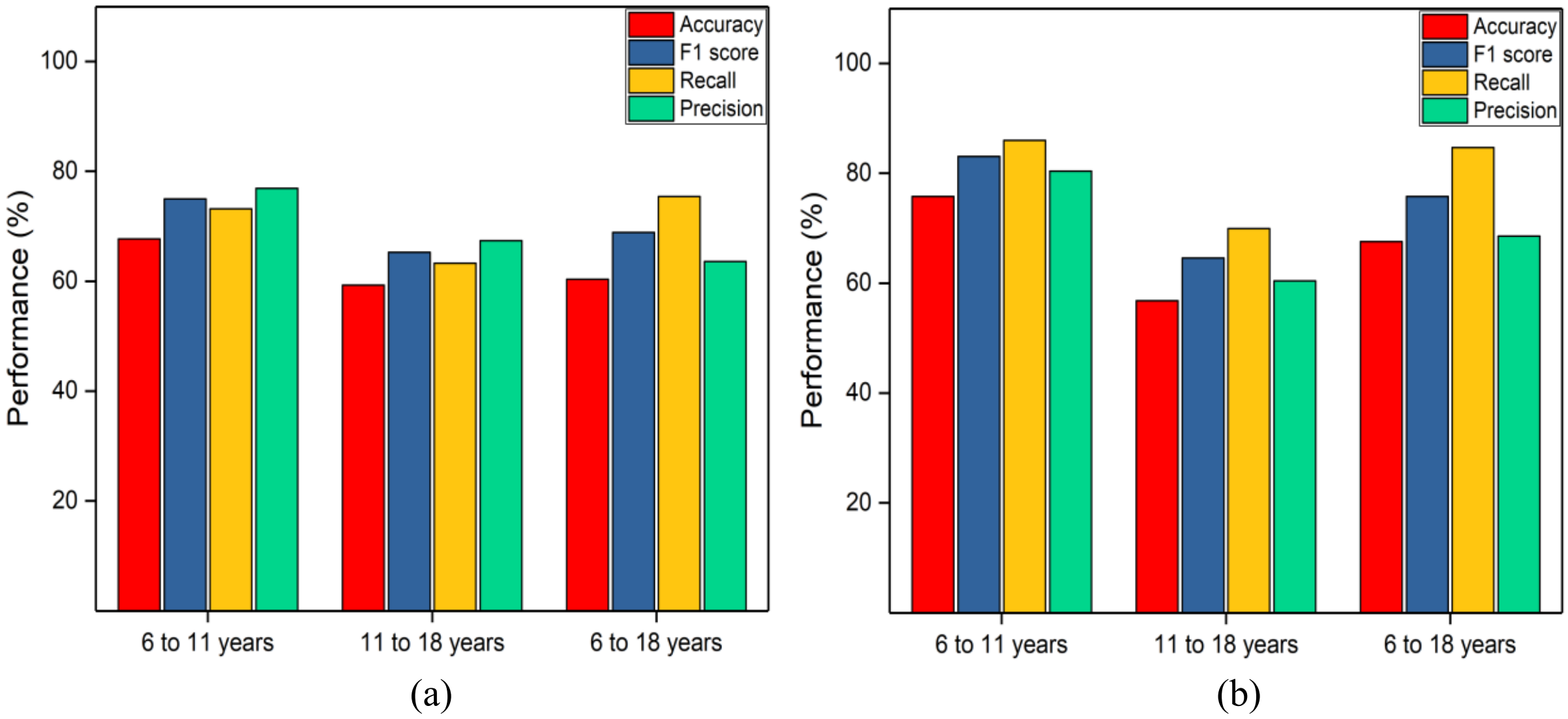

**Fig. 3** Performance metrics of the classifier on different age groups using (a) MF and (b) MCF

We observed that out of the 592 MF, the mean curvature contributed the highest number of features to the classification model. These were followed by the thickness, volume, and area features. The fraction of MF and MCF contributed by the different brain regions in the three age groups to the classifier is given in Table 2. We further analyzed the anatomical locations of the top 100 features from the different lobes for the MCF. Figure 4 represents the connectome representation of the MCF in the three different age groups. Overall, we observed that in all cases, the highest number of features were obtained from the frontal lobe.

**Table 2** Fraction of MF and MCF contributed by the different brain regions to the classifier.

| Age group (years) | Brain areas from MF (Percentage of features contributed) | Brain areas from MCF (Percentage of features contributed) |
|---|---|---|
| 6 to 11 | Frontal (48.31%)<br>Parietal (14.60%)<br>Temporal (12.35%)<br>Occipital (12.35%)<br>Insula (5.61%)<br>Occipitotemporal (3.37%)<br>Limbic (3.37%) | Frontal (28.56%)<br>Occipital (15.97%)<br>Parietal (13.84%)<br>Occipitotemporal (12.69%)<br>Temporal (11.25%)<br>Insula (10.10%)<br>Limbic (7.55) |
| 11 to 18 | Frontal (28.57%)<br>Limbic (28.57%)<br>Occipital (19.04%) | Frontal (29.16%)<br>Occipital (15.32%)<br>Temporal (14.58%) |

| Age group (years) | Brain areas from MF (Percentage of features contributed) | Brain areas from MCF (Percentage of features contributed) |
|---|---|---|
| | Insula (9.52%)<br>Temporal (4.76%)<br>Occipitotemporal (4.76%)<br>Parietal (4.76%) | Parietal (12.64%)<br>Occipitotemporal (12.05%)<br>Limbic (8.77%)<br>Insula (7.44%) |
| 6 to 18 | Frontal (30.18%)<br>Temporal (20.75%)<br>Occipital (18.86%)<br>Insula (16.98%)<br>Parietal (5.66%)<br>Occipitotemporal (3.77%)<br>Limbic (3.77%) | Frontal (22.47%)<br>Occipitotemporal (15.92%)<br>Temporal (14.43%)<br>Occipital (14.28%)<br>Parietal (13.09%)<br>Insula (11.60%)<br>Limbic (8.18%) |

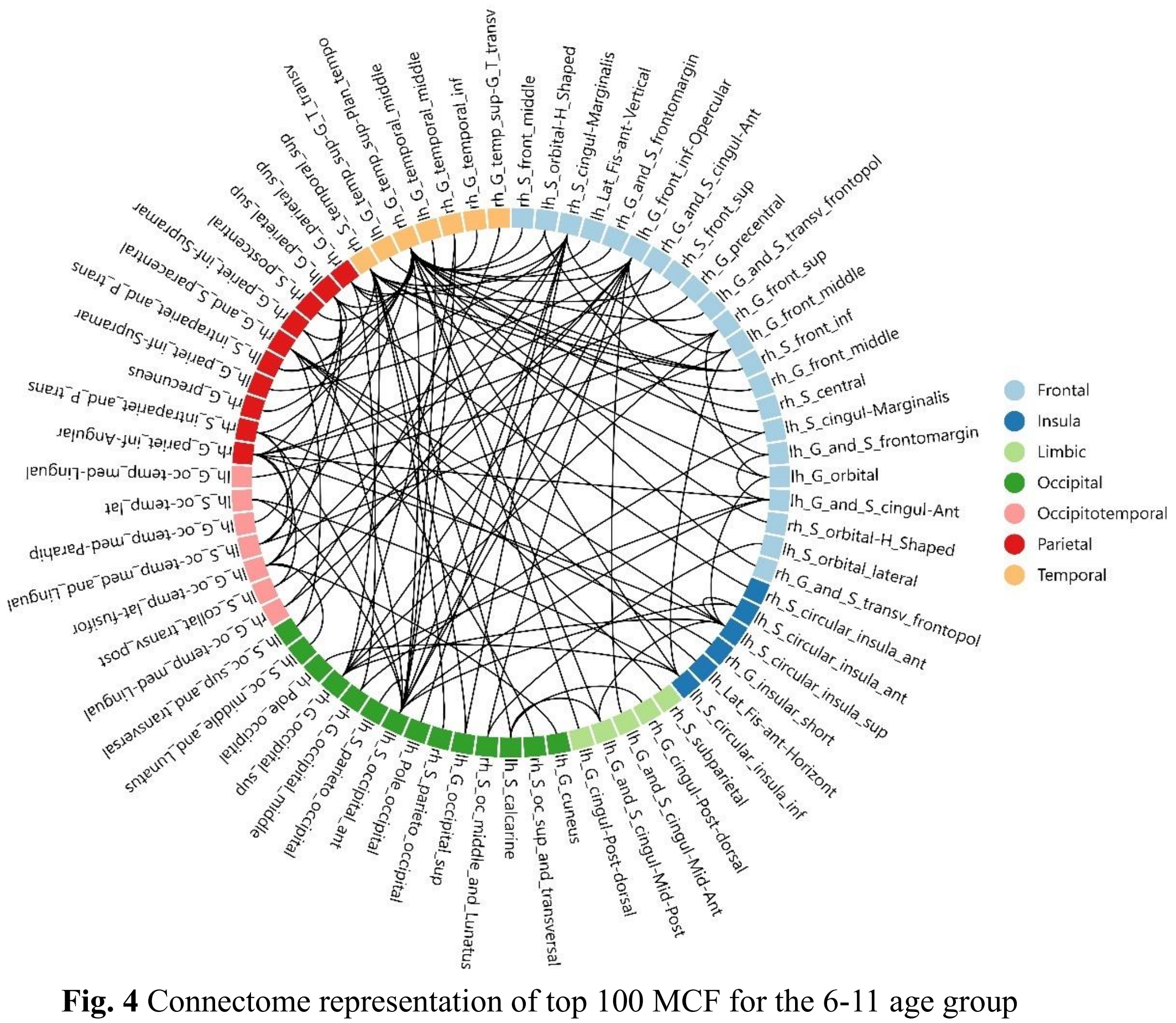

**Fig. 4** Connectome representation of top 100 MCF for the 6-11 age group

## 4. Discussion

We used MF and MCF to train an RF classifier to compare the effect of age stratification on ASD diagnosis. We achieved the highest classification accuracy in the 6-11 age group for both MF and MCF, followed by the 6-18 and 11-18 age groups. This may be due to the accelerated growth of the brain during early development in ASD subjects in contrast to the refined growth in TD subjects [18]. Studies have proved that children with ASD were well discriminated from the TD compared to the adolescent age group [19], [20]. Although the results of our classifier are low with respect to these studies, a direct comparison is unreliable as they have studies used different modalities [19], [20] and number of samples [20]. On the other hand, studies have reported higher accuracy in adults (> 18) than the adolescents (< 18) [21], [22]. However, the combined accuracy is less than the age-stratified groups, supporting our results. Moreover, our study performed better compared to the other study that used sMRI [22] Additionally, the overall classification performance of the MCF was better than the MF. It reveals that MCF can offer significant insights into the interregional morphological relationships between different brain regions. Similar results were reported in earlier studies for attention-deficit hyperactivity disorder [7][15]. Our results also suggest that features from the frontal contribute significantly to the diagnostic classification of ASD. Past studies have shown that the frontal lobe undergoes a significant morphological change in ASD [18], [23], and the features from the frontal lobe could be an effective marker in classifying ASD.

**Table 3** Comparison with existing age-specific studies

| **Study** | **Database** | **Number of Subjects** | **Age groups (years)** | **Modality** | **Features** | **Classifier** | **Performance** |
|---|---|---|---|---|---|---|---|
| **Our Study** | ABIDE I & II | 313 ASD and 397 TD | 6-11, 11-18, 6-18 | sMRI | MF and MCF | RF | 75.8%, 56.8%, 67.6% |
| [19] | ABIDE | 816 | 5-10, 10-15, 15-20, 20-30, >30 | fMRI | FC | SVM | 86%, 69% 78% 80% 95% |
| [20] | ABIDE I & II | 127 ASD and 130 TD | < 11, 11-18, > 18, Mixed, All | fMRI | PC | CVC, FT, LDA, SGD, Lib Linear | 95.23%, 78.57%, 83.33% 83.33% 69.04% |
| [21] | ABIDE | 505 ASD and 530 TD | < 18, > 18 | fMRI | SFBDM | SVM | MA- 78.6% MAD- 85.4% FA- 86.7% FAD-95% |
| [22] | ABIDE | 449 ASD and 451 TD | < 18, > 18, All | sMRI | VBM | PBL-McRBFN | 61.49% 70.41 % 59.73% |

**fMRI:** Functional Magnetic Resonance Imaging, **FC:** Functional Connectivity, **SVM:** Suppor Vector Machine, **PC:** Partial Correlation, **CVC:** Classification Via Clustering, **FT:** Functional Trees With Logistic Regression Functions, **LDA:** Linear Discriminant Analysis, **SGD:** Stochastic Gradient Descent, **Lib Linear:** Library for Large Linear Classification, **SFBDM:** Spatial Feature Based Detection Method, **MA:** Male Adolescent, **MAD:** Male Adult, **FA:** Female Adolescent, **FAD:** Female Adult, **VBM:** Voxel-Based Morphometry, **PBL-McRBFN:** Projection Based Learning Metacognitive Radial Basis Function Network Classifier.

## 5. Limitations and Future Scope

The results of our study suggest that an age-stratified approach using MCF could be an effective method to discriminate ASD. But our study has certain limitations. Our focus was confined to individuals within the 6 to 18 age group, primarily due to the constrained availability of samples from other age ranges. Additionally, we chose not to use longitudinal or gender-based categorizations, largely due to the constraints of the limited availability of relevant data. Our method of employing the Euclidean distance exclusively for MCF computation, along with employing a singular classifier for the classification task, could potentially have introduced certain biases or oversights. It is worth noting that there's ample room for expansion and refinement in this regard. By incorporating data from a wider range of sources and databases, the scope of our study could be broadened significantly.

Future studies could use a more comprehensive approach, including subject data from other databases. This could yield richer insights and more accurate results. Furthermore, using a variety of machine learning and deep learning algorithms, along with diverse connectivity metrics, could possibly improve the classification model. Therefore, our study sets the foundation for a deeper and stronger grasp of distinguishing ASD, which could lead to better accuracy and wider use.

## 6. Conclusion

In this study, we highlight the effect of age stratification in the classification of ASD subjects using various MF and MCF. We computed the MF and MCF for subjects in three age groups, 6 to 11, 11 to 18, and 6 to 18, and compared their performance using an RF classifier. Our results suggest that the 6 to 11 age group with the MCF performed the best with an accuracy, F1 score, recall, and precision of 75.8%, 83.1%, 86%, and 80.4%, respectively. We also found that out of the MF, the mean curvature was the best-contributing feature, and overall, the frontal lobe contributed the highest number of features for both the MF and MCF. The findings of our study suggest that an age-stratified approach, along with the MCF, could be an effective method to discriminate ASD.

## Statements and Declarations

### Funding

The authors declare that no funds, grants, or other support were received during the preparation of this manuscript.

### Competing Interests

The authors have no relevant financial or non-financial interests to disclose.

### Author Contributions

All authors contributed to the study's conception and design. Material preparation, data collection, and analysis were performed by Gokul Manoj and Jac Fredo. The first draft of the manuscript was written by Gokul Manoj and Jac Fredo, and all authors commented on previous versions of the manuscript. All authors read and approved the final manuscript.